\documentclass[preprint,10pt,preprintnumbers]{elsarticle}
\usepackage{graphicx,amsmath,nccmath,xcolor,float}
\usepackage{etoolbox}
\usepackage[breaklinks=true]{hyperref}
\usepackage[utf8]{inputenc}
\usepackage{multicol}
\usepackage{cuted}
\usepackage[a4paper, total={18cm, 21cm}]{geometry}
\newcommand{\Graph}[2]{\vcenter{\hbox{\includegraphics[scale=#1]{#2}}}}
\definecolor{darkblue}{rgb}{0.0,0.0,0.7}
\definecolor{darkgreen}{rgb}{0.1,0.35,0.1}
\newcommand{\CS}{\textcolor{darkblue}}
\newcommand{\FF}{}
\newcommand{\myldots}{...}
\newcommand{\mint}[1]{\scalebox{0.85}{#1}}
\makeatletter
\patchcmd{\ps@pprintTitle}{\footnotesize\itshape
       Preprint submitted to \ifx\@journal\@empty Elsevier
       \else\@journal\fi\hfill\today}{MSUHEP-21-006}{}{}
\makeatother

\graphicspath{ {fig/}{../fig} }

\allowdisplaybreaks[3]
\begin{document}
\begin{frontmatter}

\title{Four-loop collinear anomalous dimensions in QCD and $\mathcal{N} = 4$ super Yang-Mills}
\author[a]{Bakul Agarwal}\ead{agarwalb@msu.edu}
\author[a]{Andreas von Manteuffel}\ead{vmante@msu.edu}
\author[b]{Erik Panzer}\ead{erik.panzer@maths.ox.ac.uk}
\author[a]{Robert M.~Schabinger}\ead{schabing@msu.edu}
\address[a]{Department of Physics and Astronomy, Michigan State
University, East Lansing, Michigan 48824, USA}
\address[b]{Mathematical Institute, University of Oxford, OX2 6GG, Oxford, UK}

\begin{abstract}
We calculate the collinear anomalous dimensions in massless four-loop QCD and $\mathcal{N} = 4$ supersymmetric Yang-Mills theory from the infrared poles of vertex form factors.
We give very precise numerical approximations and a conjecture for the complete analytic results in both models we consider.
\end{abstract}
\end{frontmatter}

\begin{multicols}{2}

\section{Introduction}
Over the last several years, significant attention has been given to the calculation of cusp and collinear anomalous dimensions in massless perturbation theory.
The light-like \emph{cusp anomalous dimensions} \cite{Korchemsky:1987wg} enter the leading infrared poles of massless scattering amplitudes and have recently been calculated to four-loop order both in Quantum Chromodynamics (QCD) and $\mathcal{N} = 4$ supersymmetric Yang-Mills theory ($\mathcal{N} = 4$ SYM)~\cite{Grozin:2015kna,Henn:2016men,Ruijl:2016pkm,vonManteuffel:2016xki,Lee:2016ixa,Lee:2017mip,Moch:2017uml,Moch:2018wjh,Grozin:2018vdn,Lee:2019zop,Henn:2019rmi,Bruser:2019auj,vonManteuffel:2019wbj,Henn:2019swt,Huber:2019fxe,vonManteuffel:2020vjv}.
The \emph{collinear anomalous dimensions} enter the subleading infrared poles and can be extracted from the $1/\epsilon$ poles of vertex {\it form factors} \cite{Mueller:1979ih,Collins:1980ih,Sen:1981sd,Magnea:1990zb,Sterman:2002qn,Ravindran:2004mb,Dixon:2008gr}.
Partial results are available at four-loop order both in QCD \cite{Henn:2016men,Lee:2016ixa,Lee:2017mip,Lee:2019zop,Moch:2017uml,Das:2019btv,Das:2020adl,vonManteuffel:2020vjv} and $\mathcal{N} = 4$ supersymmetric Yang-Mills theory ($\mathcal{N} = 4$ SYM) \cite{Cachazo:2007ad,Boels:2017ftb,Dixon:2017nat}.
In this article, we consider the analytically unknown four-loop contributions to the collinear anomalous dimensions.

In QCD, the basic quark and gluon form factors are the normalized amplitudes for, respectively, a virtual photon decaying into a pair of massless quarks, $\gamma^\ast(q)\rightarrow q(p_1)\bar{q}(p_2)$, and a Higgs boson decaying into two gluons in the limit of infinite top quark mass, $h(q)\rightarrow g(p_1)g(p_2)$, whereas in $\mathcal{N} = 4$ SYM, the Sudakov form factor is the normalized amplitude
\begin{align}
\label{eq:def}
\FF{\bar{\mathcal{F}}^{\mathcal{N} = 4}} = \frac{1}{N} \int \!{\rm d}^4 x \,\,e^{-i\, q \cdot x}
\notag\\
\langle \phi_{1 2}^a(p_1) &\phi_{1 2}^b(p_2) |
\left[\phi_{3 4}^c \phi_{3 4}^c\right]\!(x) \,|0\rangle,
\end{align}
where field superscripts denote adjoint $SU(N_c)$ color indices, field subscripts denote $SU(4)_R$ indices, and the constant $N$ is chosen such that $\FF{\bar{\mathcal{F}}^{\mathcal{N} = 4}}$ is one at leading order.

In QCD, the perturbative expansion of the bare form factors is
\begin{equation}
\label{eq:expbareg}
    \FF{\bar{\mathcal{F}}_{\rm bare}^r} =
1 + \sum_{L = 1}^\infty
        \left(\frac{\alpha_s^{\rm bare}}{4\pi}\right)^L 
        \left(\frac{4\pi \mu_\epsilon^2}{-q^2 e^{\gamma_E}}\right)^{L \epsilon}
        \FF{\bar{\mathcal{F}}_L^r(\epsilon)}\,,
\end{equation}
where $r = q$ or $g$ for quarks or gluons, $\alpha_s^{\rm bare}$ is the bare coupling, $q^2 = (p_1 + p_2)^2$ is the virtuality, $\mu_\epsilon$ is the 't\,Hooft scale, $\epsilon = (4-d)/2$ is the parameter of dimensional regularization, and $\gamma_E$ is Euler's constant. In $\mathcal{N} = 4$ SYM, the expansion is
\begin{equation}
 \FF{\bar{\mathcal{F}}^{\mathcal{N} = 4}} =  1 + \sum_{L = 1}^\infty \lambda^L\left(\frac{\mu_\epsilon^2}{-q^2}\right)^{L \epsilon}\FF{\bar{\mathcal{F}}_L^{\mathcal{N} = 4}(\epsilon)},
\end{equation}
in terms of the modified bare 't\,Hooft coupling
\begin{equation}
\label{eq:tHooft}
\lambda = \frac{N_c\, g_{\scriptscriptstyle \mathcal{N} = 4}^2}{16\pi^2}\left(4\pi e^{-\gamma_E}\right)^\epsilon,
\end{equation}
and $g_{\scriptscriptstyle \mathcal{N} = 4}$ is the bare coupling of the $\mathcal{N} = 4$ SYM model.

The collinear anomalous dimensions of QCD receive the four-loop contributions
\begin{align}
\label{eq:colqgdef}
\FF{\gamma_4^r} &= G^r_4[0] - \beta_0 G^r_3[1] - \beta_1 G^r_2[1] - \beta_2 G^r_1[1]
\nonumber \\& \quad
 \vphantom{\Big(}+ \beta_0^2 G^r_2[2] + 2 \beta_0 \beta_1 G^r_1[2] - \beta_0^3 G^r_1[3] + 8 \beta_3 \delta_{g r},
\end{align}
where we follow \cite{Moch:2005id}, see also \cite{Ravindran:2006cg}. In Eq.\ \eqref{eq:colqgdef}, $G^r_L[k]$ denotes the $\epsilon^k$ coefficient of the {\it resummation function} $G^r_L(\epsilon)$ as defined in Eqs.\ (2.14)-(2.17) of \cite{Moch:2005id}, and $\beta_{L-1}$ denotes the massless QCD beta function coefficient of order $L$, see {\it e.g.}\ \cite{vanRitbergen:1997va,Czakon:2004bu} for explicit results.
In $\mathcal{N} = 4$ SYM, the absence of a running coupling implies
\begin{equation}
    \FF{\gamma_4^{\mathcal{N} = 4}} = G^{\mathcal{N} = 4}_4[0]\,.
\end{equation}
The $1/\epsilon$ poles of the form factors allow for a determination of the resummation functions and thus the collinear anomalous dimensions.

The remainder of this article is organized as follows. In Section \ref{sec:compmethod}, we describe our computational methods based on integration by parts reductions and direct integration of Feynman parametric representations; we also give results for some master integrals.
In Section \ref{sec:results}, we present analytic results for the $1/\epsilon$ poles of the four-loop form factors $\bar{\mathcal{F}}_4^q(\epsilon)$, $\bar{\mathcal{F}}_4^g(\epsilon)$, and $\bar{\mathcal{F}}_4^{\mathcal{N} = 4}(\epsilon)$, and the corresponding collinear anomalous dimensions $\gamma_4^q$, $\gamma_4^g$, and $\gamma_4^{\mathcal{N} = 4}$.
Our analytic results are expressed in terms of zeta values and a single leading-order-in-$\epsilon$ coefficient of a finite master integral, which could not be straightforwardly handled by the {\tt HyperInt} program \cite{Panzer:2014caa}.
In Section \ref{sec:numresults}, we employ a very precise numerical approximation of this integral coefficient to provide complete numerical results for all form factors and collinear anomalous dimensions.
In Section \ref{sec:pslqanalysis}, we perform a PSLQ analysis \cite{PSLQ} to lift our precise numerical data to conjectured analytic results and describe a number of plausibility arguments which support our conjecture.
Finally, in Section \ref{sec:summary}, we conclude.

\section{Computational methods}
\label{sec:compmethod}

\begin{figure*}
\begin{align*}
\Graph{.25}{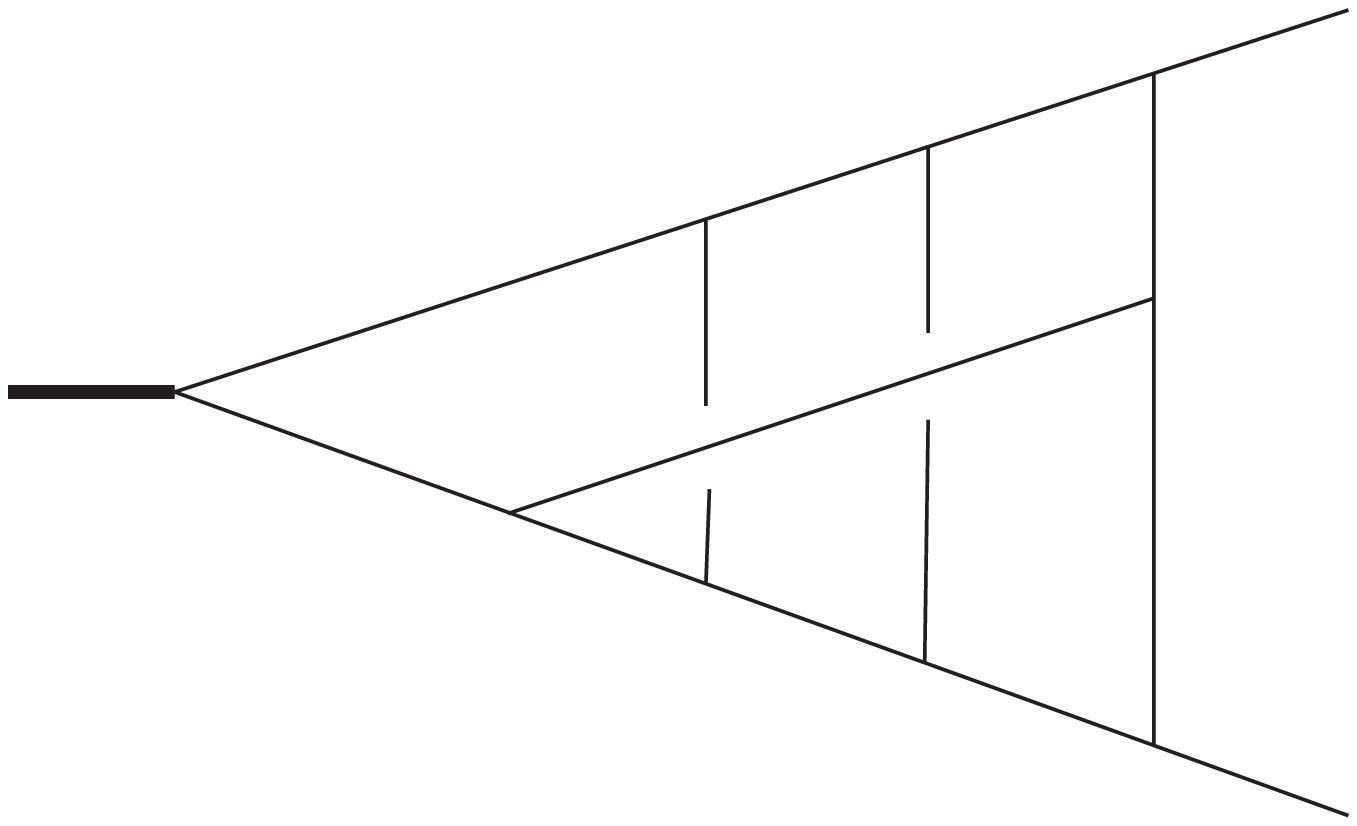} \qquad \qquad \Graph{.25}{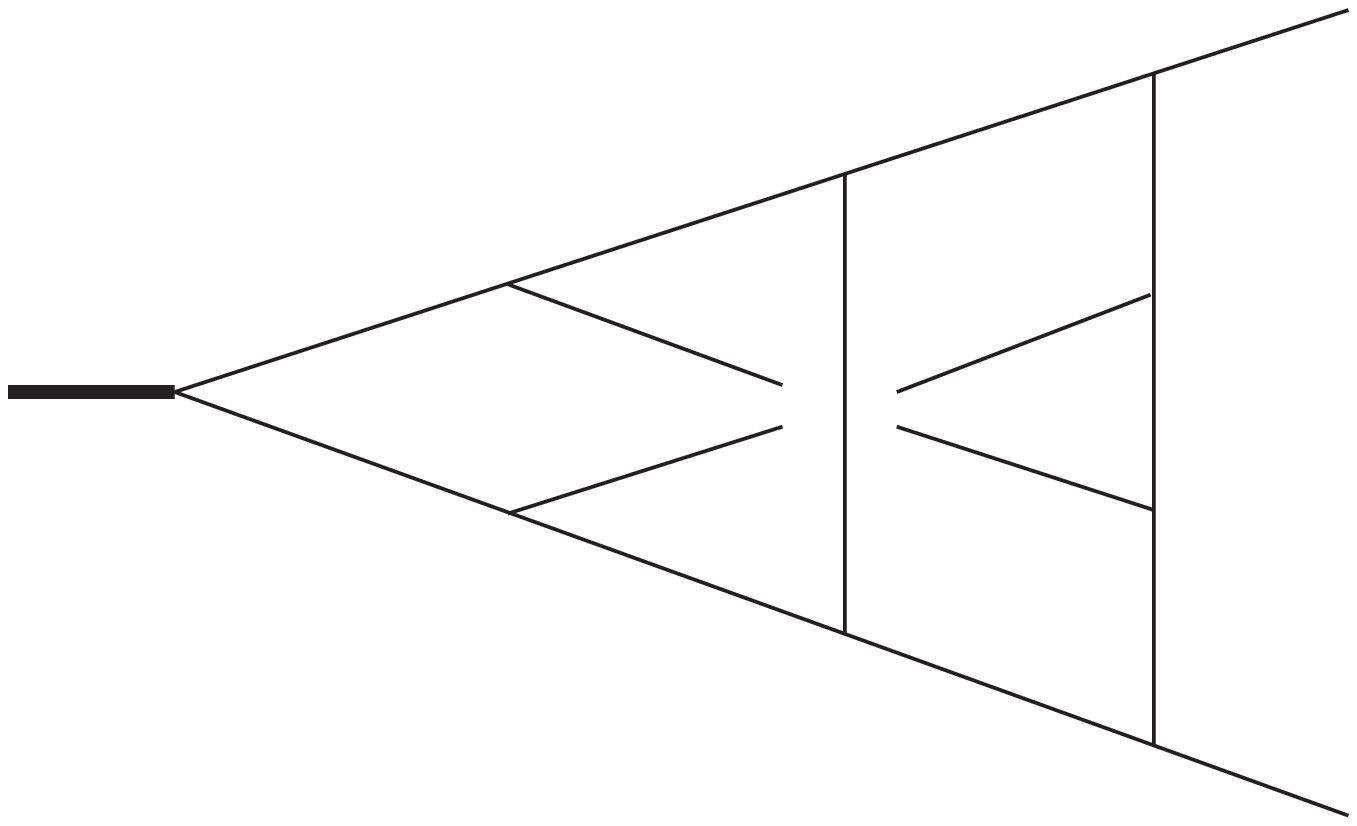}
\end{align*}
\caption{
Top-level topologies whose Feynman parametric representation is not linearly reducible and therefore not directly accessible to {\tt HyperInt} for generic integrals.
}
\label{fig:hardtopos}%
\end{figure*}

Our analytic calculation of the form factors follows that of \cite{vonManteuffel:2020vjv}, 
employing a primary integration by parts reduction \cite{Tkachov:1981wb,Chetyrkin:1981qh,Gehrmann:1999as,Laporta:2001dd,Gluza:2010ws,CabarcasDing,Schabinger:2011dz,vonManteuffel:2014ixa,vonManteuffel:2016xki,Ita:2015tya,Larsen:2015ped,Boehm:2017wjc} and a subsequent rotation \cite{Lee:2013hzt,Lee:2013mka,Bitoun:2017nre} to a judiciously-chosen basis of finite master integrals \cite{vonManteuffel:2014qoa,vonManteuffel:2015gxa,Schabinger:2018dyi,vonManteuffel:2019gpr}, computed with a private implementation, \texttt{Finred}.
We make heavy use of {\tt HyperInt} for the analytic evaluation of the master integrals.
However, for the two topologies shown in \autoref{fig:hardtopos} the generic Feynman parametric representation that we use is not {\it linearly reducible} \cite{Brown:2008um,Brown:2009ta}, that is, it cannot be directly integrated with the algorithm \cite{Panzer:2014caa}.
We have been able to find linearly reducible integrands only for the \textit{leading-order} $\epsilon$-expansion coefficients of \textit{specific} integrals in these topologies as will be explained in the following.
 However, we do not know whether generic integrals in these topologies can be rendered linearly reducible to all orders by changing variables.

For the topology on the left-hand side of \autoref{fig:hardtopos}, we found a basis of two finite integrals in $d=6-2\epsilon$ dimensions, which appear for the first time at the level of the $1/\epsilon$ poles {\it and} which each have 15 as the sum of their propagator exponents. Due to this choice, the exponent of the Symanzik polynomial $\mathcal{U}$ vanishes at zeroth order in the $\epsilon$ expansion of these integrals -- the {\it only} order which we need to obtain results for $\gamma_4^q$, $\gamma_4^g$, and $\gamma_4^{\mathcal{N} = 4}$. The remaining polynomial $\mathcal{F}$, it turns out, is by itself linearly reducible for this topology, therefore allowing for a straightforward application of {\tt HyperInt}.

For the topology on the right-hand side of \autoref{fig:hardtopos}, the situation is more complicated, despite the fact that we find a change of variables which renders the $\mathcal{F}$ polynomial linearly reducible in this case as well.
First, we count four master integrals for the topology in $d$ dimensions, but were able to choose a basis such that only three of them contribute to the $1/\epsilon$ poles of the form factors.
Unfortunately, we did not find a suitable basis of integrals for the $1/\epsilon$ pole such that all of them are independent of the $\mathcal{U}$ polynomial, {\it e.g.}\ by choosing 15 for the sum of the propagator exponents for all integrals in $d=6-2\epsilon$ dimensions.

Instead, the best-case scenario seems to allow for a straightforward treatment of two out of three finite integrals only; the remaining finite integral which contributes to the $1/\epsilon$ poles can be chosen to have 13, the minimal number, as the sum of its propagator exponents. Unfortunately, an additional complication arises due to the fact that making such a choice of finite integrals requires the computation of some finite integrals in subtopologies to higher orders in the $\epsilon$ expansion than would have been necessary in the finite integral basis of \cite{vonManteuffel:2020vjv}, {\it i.e.} one constructed to be compatible with a basis of uniform weight as suggested in \cite{Schabinger:2018dyi}.

Our auxiliary results include, for example, the $\mathcal{O}\left(\epsilon^2\right)$ term of
\begin{align}
\label{eq:C_10_32172}
 &\hspace{-1ex}\overset{6-2\epsilon}{\Graph{.2}{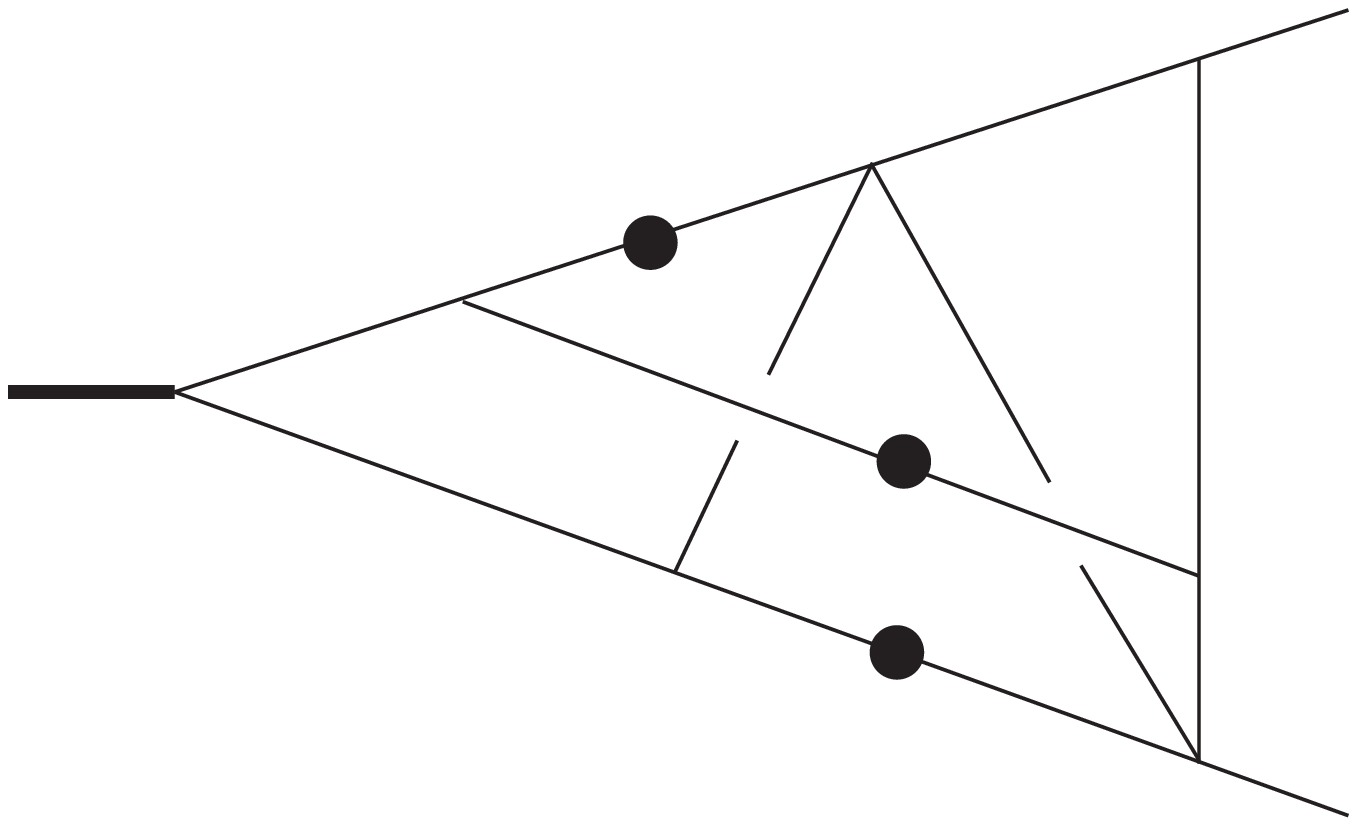}}  = -\mfrac{53}{6}\zeta_3^2-\mfrac{1579}{630}\zeta_2^3+\mfrac{535}{36}\zeta_5
+\mfrac{5}{6}\zeta_3 \zeta_2
\nonumber\\ &
 -\mfrac{17}{12} \zeta_2^2+\mfrac{121}{18}\zeta_3+\mfrac{8}{9} \zeta_2
 +\textcolor{darkgreen}{\epsilon} \Big(\!\!-\mfrac{1033}{6} \zeta_7 
 -\mfrac{53}{3}\zeta_5\zeta_2+\mfrac{21}{5} \zeta_3\zeta _2^2
\nonumber\\ &
 +\mfrac{139}{18} \zeta_3^2+\mfrac{1609}{126} \zeta_2^3+\mfrac{6235}{108} \zeta_5
 -\mfrac{145}{18} \zeta_3 \zeta_2-\mfrac{49}{12} \zeta_2^2 +\mfrac{1478}{27} \zeta_3
\nonumber\\ &
 +\mfrac{565}{54} \zeta_2\!\Big)
 +\textcolor{darkgreen}{\epsilon^2} \Big(\mfrac{10403}{30} \zeta_{5,3} +\mfrac{2887}{6} \zeta_5\zeta_3
 +\mfrac{307}{2} \zeta_3^2 \zeta_2 
\nonumber \\ &
 -\mfrac{247041}{1000} \zeta_2^4 +\mfrac{78607}{144}\zeta_7+\mfrac{110}{9} \zeta_5 \zeta_2+\mfrac{2461}{180} \zeta_3 \zeta_2^2 -\mfrac{18467}{216} \zeta_3^2
\nonumber \\ &
 +\mfrac{30287}{2268} \zeta_2^3 +\mfrac{79117}{324} \zeta_5-\mfrac{30455}{108}\zeta_3 \zeta_2
 -\mfrac{1495}{216} \zeta_2^2 +\mfrac{112325}{324} \zeta_3
\nonumber\\ &
 +\mfrac{6854}{81} \zeta_2\!\Big) + \textcolor{darkgreen}{\mathcal{O}\!\left(\epsilon^3\right)}.
\end{align}
To obtain this and other results, we ran {\tt HyperInt} in a highly parallelized setup, accumulating several CPU years in total.\footnote{We use the convention
$
\zeta_{5,3}=\sum_{m = 1}^\infty \frac{1}{m^{5}} \sum_{n = 1}^{m-1}\frac{1}{n^{3}}
\approx 0.0377076729848\myldots
$
and the normalization conventions of \cite{vonManteuffel:2015gxa} for Eq.\ \eqref{eq:C_10_32172}.}
For the most complicated topology, we determined two integrals with a propagator exponent sum of 15 analytically,
\begin{align}
&\hspace{-1ex}\overset{6-2\epsilon}{\Graph{.2}{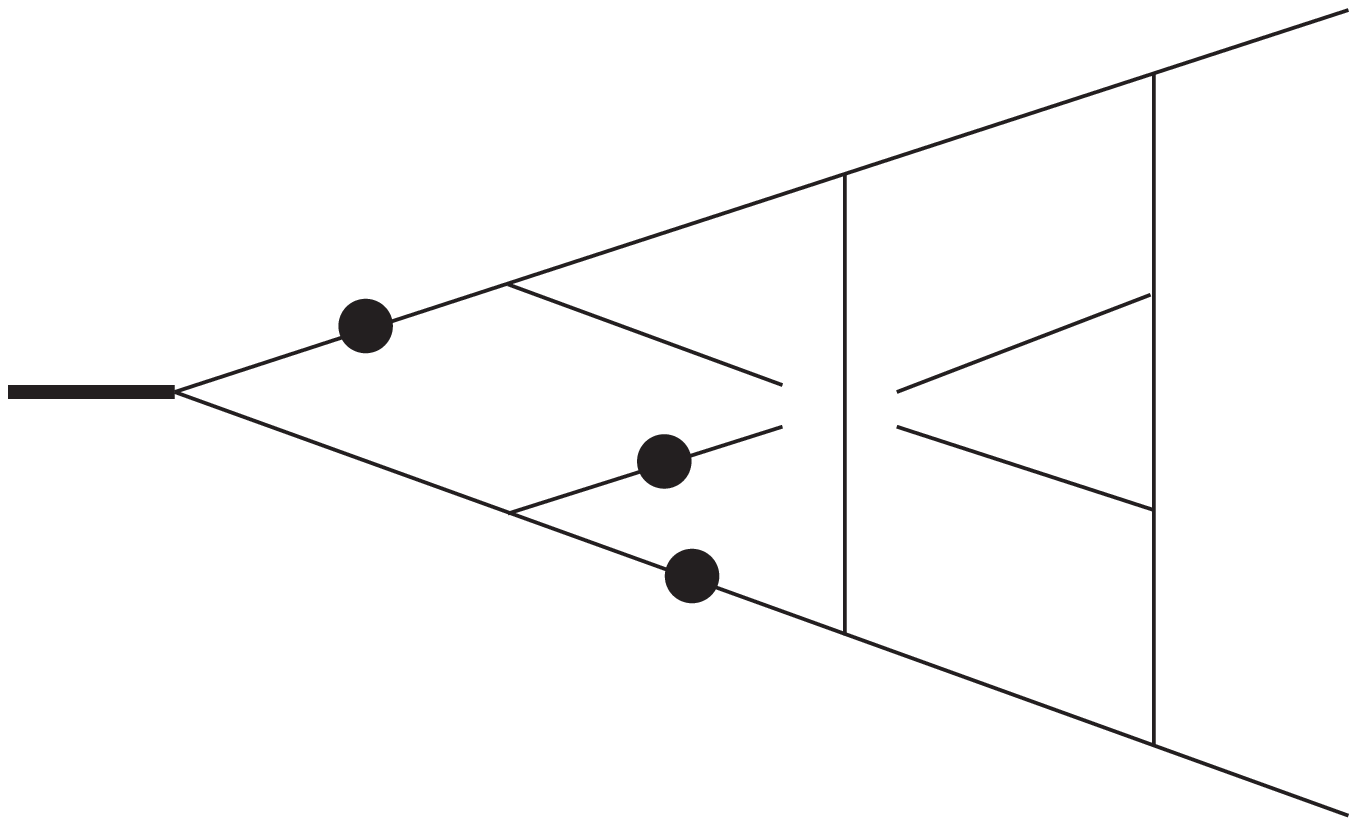}}  = -\mfrac{4221}{16} \zeta _7+\mfrac{159}{2} \zeta _5 \zeta _2 +\mfrac{25}{2} \zeta _3 \zeta _2^2 \notag\\
&\quad -\mint{14} \zeta _3^2 -\mfrac{631}{70} \zeta_2^3 +\mfrac{535}{2} \zeta_5-\mint{69} \zeta_3 \zeta_2+\textcolor{darkgreen}{\mathcal{O}\left(\epsilon\right)},\\[2ex]
&\hspace{-1ex}\overset{6-2\epsilon}{\Graph{.2}{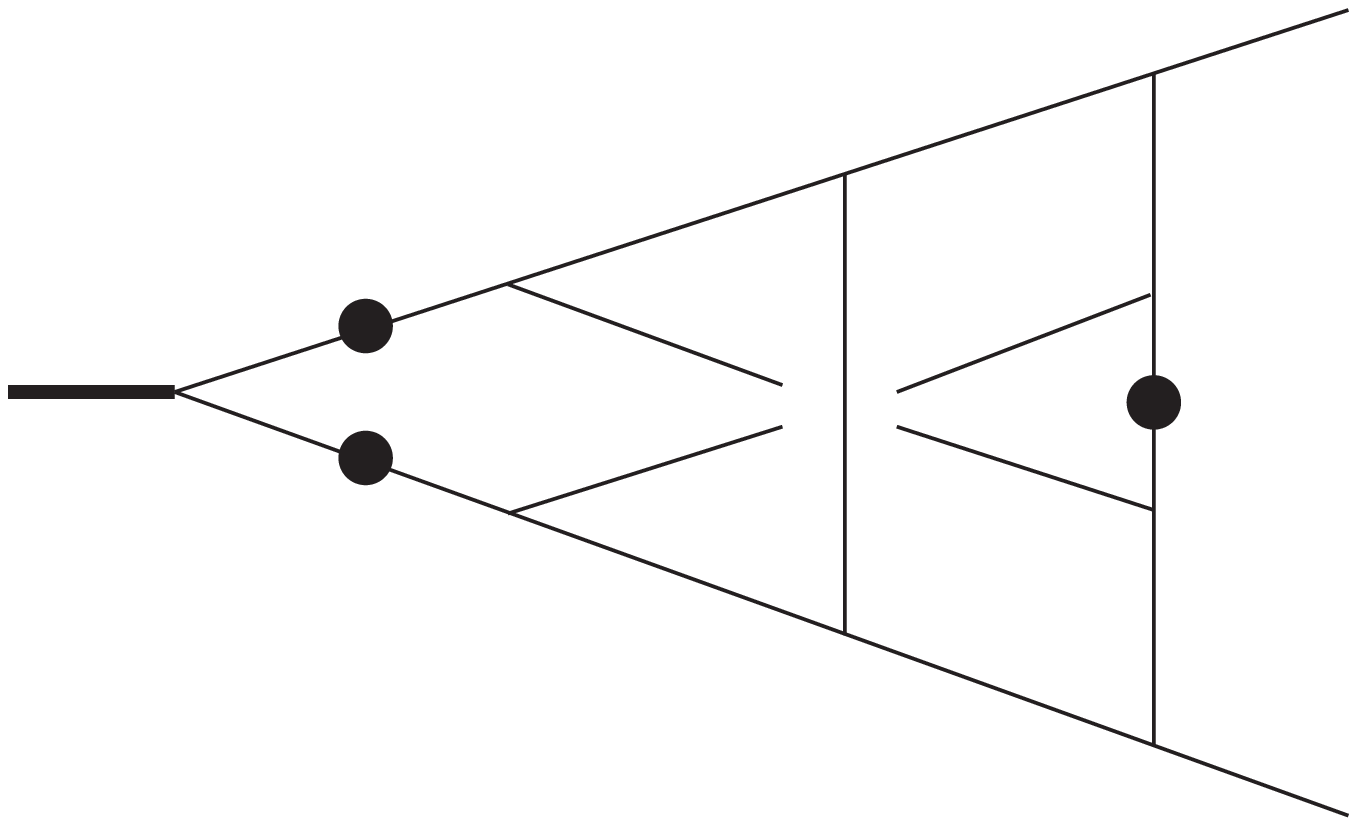}}  = \mfrac{252}{5} \zeta_{5,3}+\mint{195} \zeta_5\zeta_3-\mint{18} \zeta_3^2 \zeta_2 \notag\\
&\quad -\mfrac{202807}{10500} \zeta_2^4 -\mfrac{959}{8}\zeta_7 +\mint{50} \zeta_5 \zeta_2-\mfrac{51}{5} \zeta_3 \zeta_2^2+\textcolor{darkgreen}{\mathcal{O}\left(\epsilon\right)},
\\
\intertext{leaving only}
&\hspace{-1ex}\overset{6-2\epsilon}{\Graph{.2}{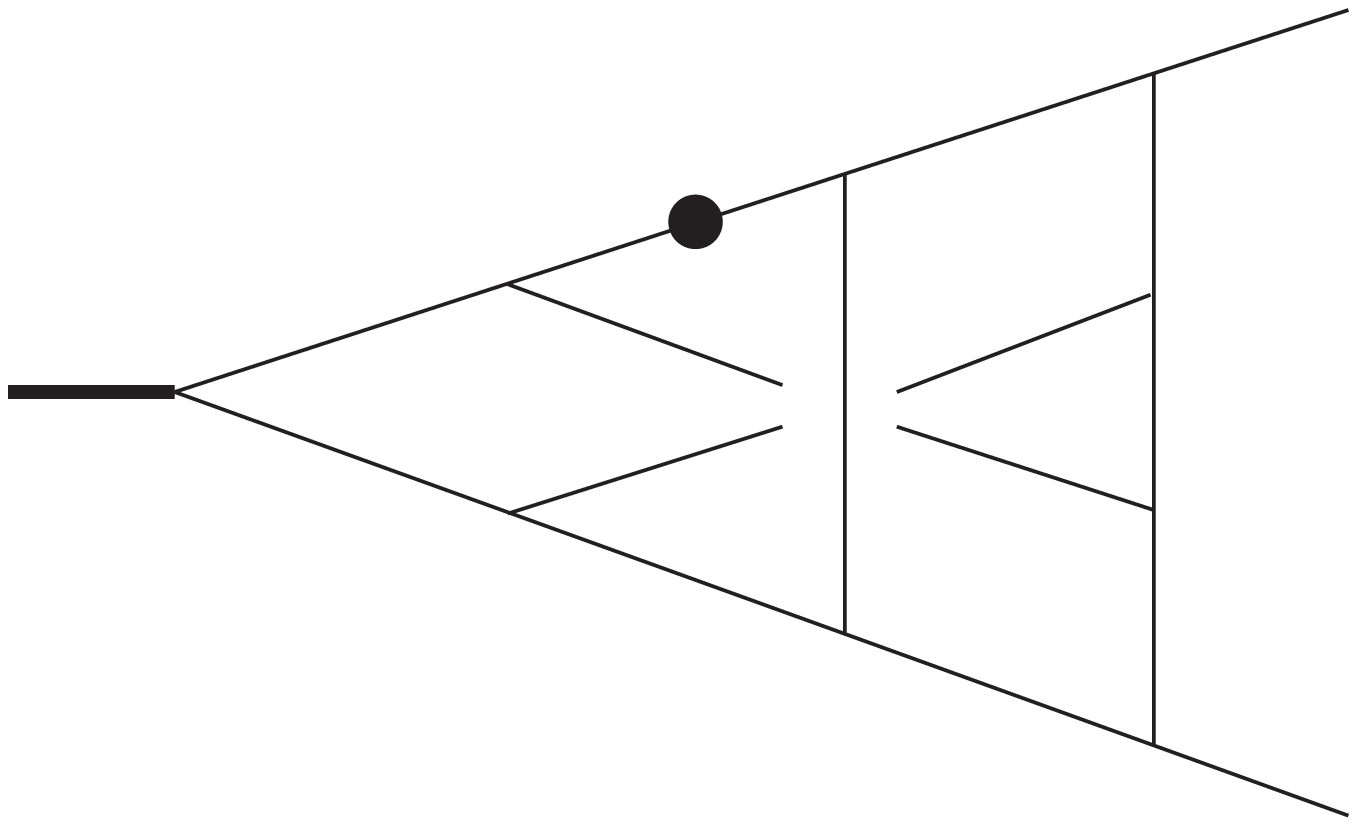}} \equiv \mathcal{H} + \textcolor{darkgreen}{\mathcal{O}\left(\epsilon\right)}.
\end{align}

We were able to evaluate $\mathcal{H}$ to high precision numerically by running the {\tt pySecDec} program \cite{Borowka:2017idc} for several months in a distributed manner on high-performance GPUs,
\begin{equation}
\label{eq:num}
    \mathcal{H} \approx 
    - \mint{0.7015802723647} \pm \mint{6.98} \cdot \mint{10}^{-11} \,,
\end{equation}
where, with some foresight, we have appropriately rounded our numerical result and the provided estimate of its statistical uncertainty. While we expect our numerical results to suffice for phenomenological applications, we will see later that the precision of our results even allows us to put forth a plausible conjecture for the analytical form of $\mathcal{H}$ and, therefore, of the four-loop collinear anomalous dimensions.

In contrast to the case of QCD, the reduced integrand for the Sudakov form factor of the $\mathcal{N} = 4$ SYM model is known as a simple linear combination of (conjecturally) uniform weight Feynman integrals defined in \cite{Boels:2017ftb} and evaluated through to weight six in \cite{Huber:2019fxe}:
\begin{align}
\label{eq:neq4integrand}
 &\FF{\bar{\mathcal{F}}_4^{\mathcal{N} = 4}(\epsilon)} = \mint{2} \Bigg[\mint{8} I_{\mathrm{p}, 1}^{(1)}+\mint{2} I_{\mathrm{p}, 2}^{(2)}-\mint{2} I_{\mathrm{p}, 3}^{(3)}+\mint{2} I_{\mathrm{p}, 4}^{(4)}+\mfrac{1}{2} I_{\mathrm{p}, 5}^{(5)} \nonumber \\
 &\quad +\mint{2} I_{\mathrm{p}, 6}^{(6)}+\mint{4} I_{\mathrm{p}, 7}^{(7)}+\mint{2} I_{\mathrm{p}, 8}^{(9)}
 -\mint{2} I_{\mathrm{p}, 9}^{(10)}+ I_{\mathrm{p}, 10}^{(12)}+ I_{\mathrm{p}, 11}^{(12)}\nonumber \\
 &\quad +\mint{2} I_{\mathrm{p}, 12}^{(13)} +\mint{2} I_{\mathrm{p}, 13}^{(14)}-\mint{2} I_{\mathrm{p}, 14}^{(17)}+\mint{2} I_{\mathrm{p}, 15}^{(17)}-\mint{2} I_{\mathrm{p}, 16}^{(19)} + I_{\mathrm{p}, 17}^{(19)}\nonumber \\
 &\quad + I_{\mathrm{p}, 18}^{(21)} +\mfrac{1}{2} I_{\mathrm{p}, 19}^{(25)}+\mint{2} I_{\mathrm{p}, 20}^{(30)}+\mint{2} I_{\mathrm{p}, 21}^{(13)}+\mint{4} I_{\mathrm{p}, 22}^{(14)}-\mint{2} I_{\mathrm{p}, 23}^{(14)} \nonumber \\
 &\quad - I_{\mathrm{p}, 24}^{(14)} +\mint{4} I_{\mathrm{p}, 25}^{(17)}- I_{\mathrm{p}, 26}^{(17)}-\mint{2} I_{\mathrm{p}, 27}^{(17)}-\mint{2} I_{\mathrm{p}, 28}^{(17)}- I_{\mathrm{p}, 29}^{(19)} \nonumber \\
 &\quad - I_{\mathrm{p}, 30}^{(19)} + I_{\mathrm{p}, 31}^{(19)}-\mfrac{1}{2} I_{\mathrm{p}, 32}^{(30)}\Bigg] +\mint{48} \CS{\frac{1}{N_c^2}}\Bigg[ \mfrac{1}{2} I_{1}^{(21)}+\mfrac{1}{2} I_{2}^{(22)} \nonumber \\
 &\quad +\mfrac{1}{2} I_{3}^{(23)} - I_{4}^{(24)}+\mfrac{1}{4} I_{5}^{(25)}-\mfrac{1}{4} I_{6}^{(26)}-\mfrac{1}{4} I_{7}^{(26)} +2 I_{8}^{(27)}\nonumber \\
 &\quad +I_{9}^{(28)} +\mint{4} I_{10}^{(29)}+ I_{11}^{(30)}+ I_{12}^{(27)}-\mfrac{1}{2} I_{13}^{(28)}+ I_{14}^{(29)}\nonumber \\
 &\quad + I_{15}^{(29)} + I_{16}^{(30)}+ I_{17}^{(30)}+ I_{18}^{(30)}+ I_{19}^{(22)}+ I_{20}^{(22)}- I_{21}^{(24)}\nonumber \\
 &\quad +\mfrac{1}{4} I_{22}^{(24)} +\mfrac{1}{2} I_{23}^{(28)}\Bigg].
\end{align}
$\mathcal{H}$ enters both $I_{6}^{(26)}$ and $I_{7}^{(26)}$ in the non-planar-color part of Eq.\ \eqref{eq:neq4integrand}, allowing for an over-determination of the lower-weight power products of zeta values which enter $\mathcal{H}$. Let us emphasize that a rotation to a basis of finite integrals is still of paramount importance for our $\mathcal{N} = 4$ SYM calculation, but, due to the uniform weight property, it is convenient to organize the calculation of the $\mathcal{N} = 4$ Sudakov form factor in terms of the master integrals entering Eq.\ \eqref{eq:neq4integrand}.

Note that, while the most complicated, non-linearly reducible integral topologies appeared in all four-loop form factors we calculated through to weight seven, the calculation of the QCD master integrals to sufficiently high orders in $\epsilon$ was harder overall, because it involved a significant number of computationally challenging, non-planar integral topologies that do not appear in the four-loop $\mathcal{N} = 4$ SYM Sudakov form factor.

\section{Analytical results}
\label{sec:results}
In this section, we present results for the $1/\epsilon$ poles of the four-loop form factors we consider, $\bar{\mathcal{F}}_4^q(\epsilon)$, $\bar{\mathcal{F}}_4^g(\epsilon)$, and $\bar{\mathcal{F}}_4^{\mathcal{N} = 4}(\epsilon)$, and the corresponding four-loop collinear anomalous dimensions, $\gamma_4^q$, $\gamma_4^g$, and $\gamma_4^{\mathcal{N} = 4}$, as a function of $\mathcal{H}$.
Here, we omit the known $N_f$-dependent terms~\cite{vonManteuffel:2020vjv} involving one, two or three closed fermion loops.
For the $1/\epsilon$ poles of the four-loop form factors, we find
\begin{align}
\label{eq:ffq}
    &\FF{\bar{\mathcal{F}}_4^q(\epsilon)\Big|_{1/\epsilon}} =  
    \CS{\frac{d_A^{abcd}d_F^{abcd}}{N_F}}\Big[-\mfrac{2481}{2} \zeta_7-\mint{328} \zeta_5 \zeta_2 +\mfrac{1092}{5} \zeta_3 \zeta_2^2 \nonumber \\
    &\;\; -\mfrac{1982}{3}\zeta_3^2 -\mfrac{30236}{315} \zeta_2^3+\mfrac{18230}{9}\zeta_5 +\mint{704} \zeta_3 \zeta_2-\mfrac{388}{15}\zeta_2^2 \nonumber \\
    &\;\; -\mfrac{5504}{9} \zeta_3+\mfrac{272}{3}\zeta_2 -\mint{24}+\mint{80} \mathcal{H}\Big] + \CS{C_A^3 C_F}\Big[\mfrac{18927}{16}\zeta_7 \nonumber \\
    &\;\; -\mfrac{31}{3}\zeta_5 \zeta_2-\mfrac{239}{10} \zeta_3 \zeta_2^2-\mfrac{44147}{36}\zeta_3^2-\mfrac{40859}{135}\zeta_2^3+\mfrac{14285}{108}\zeta_5 \nonumber \\
    &\;\; -\mfrac{226}{9}\zeta_3 \zeta_2+\mfrac{506519}{270}\zeta_2^2+\mfrac{1206563}{54}\zeta_3 -\mfrac{6051515}{648}\zeta_2 \nonumber \\
    &\;\; -\mfrac{651546401}{23328}-\mfrac{70}{3}\mathcal{H}\Big]+ \CS{C_A^2 C_F^2}\Big[-\mfrac{98381}{36}\zeta_7 +\mfrac{6571}{9}\zeta_5 \zeta_2 \nonumber\\
    &\;\; +\mfrac{56017}{135} \zeta_3 \zeta_2^2+\mfrac{654539}{81}\zeta_3^2-\mfrac{14915}{63} \zeta_2^3-\mfrac{4167568}{405}\zeta_5 \nonumber \\
    &\;\; -\mfrac{89743}{27}\zeta_3\zeta_2-\mfrac{24918629}{4860}\zeta_2^2-\mfrac{380023273}{5832}\zeta_3+\mfrac{347179283}{11664}\zeta_2 \nonumber\\
    &\;\; +\mfrac{29277646423}{419904}+\mint{60} \mathcal{H}\Big]+ \CS{C_A C_F^3}\Big[\mfrac{13149}{2} \zeta_7-\mfrac{11048}{5} \zeta_5 \zeta_2 \nonumber\\
    &\;\; -\mfrac{12008}{15} \zeta_3 \zeta_2^2-\mfrac{291697}{27} \zeta_3^2+\mfrac{712849}{945} \zeta_2^3+\mfrac{1097047}{54}\zeta_5 \nonumber\\
    &\;\; +\mfrac{101125253}{1944}\zeta_3 -\mfrac{38043757}{1296} \zeta_2+\mfrac{197249}{27} \zeta_3 \zeta_2 +\mfrac{5508667}{1620}\zeta_2^2 \nonumber\\
    &\;\; -\mfrac{6800926313}{139968} -\mint{40} \mathcal{H}\Big]+ \CS{C_F^4}\Big[-\mfrac{14162}{21} \zeta_7 +\mfrac{5792}{5}\zeta_5 \zeta_2 \nonumber \\
    &\;\; +\mfrac{6208}{9} \zeta_3 \zeta_2^2+\mfrac{14060}{9} \zeta_3^2-\mfrac{16786}{15} \zeta_2^3 -\mfrac{235816}{15}\zeta_5 \nonumber \\
    &\;\; -\mfrac{32966}{9}\zeta_3 \zeta_2 +\mfrac{479}{15} \zeta_2^2-\mfrac{87481}{18} \zeta_3 +\mfrac{26425}{3} \zeta_2 +\mfrac{94257}{8}\Big] \nonumber\\
    &\;\; + \CS{N_f\text{\small{-terms}}},
\\[2ex]
\label{eq:ffg}
    &\FF{\bar{\mathcal{F}}_4^g(\epsilon)\Big|_{1/\epsilon}} = \CS{\frac{d_A^{abcd}d_A^{abcd}}{N_A}}\Big[-\mfrac{2481}{2}\zeta_7-\mint{328} \zeta _5 \zeta _2+\mfrac{1092}{5} \zeta _3 \zeta _2^2 \nonumber\\
    &\;\;-\mfrac{1982}{3} \zeta_3^2 -\mfrac{31732}{315} \zeta _2^3 +\mfrac{17660}{9} \zeta _5+\mint{772} \zeta _3 \zeta _2-\mfrac{134}{15} \zeta_2^2 \nonumber\\
    &\;\;-\mfrac{2804}{9} \zeta _3-8 \zeta _2 -\mfrac{32}{3}+\mint{80} \mathcal{H}\Big]+ \CS{C_A^4}\Big[\mfrac{4385053}{1008} \zeta _7 \nonumber\\
    &\;\;-\mfrac{14914}{45} \zeta _5 \zeta_2+\mfrac{75677}{270} \zeta _3 \zeta _2^2-\mfrac{1098847}{324} \zeta _3^2-\mfrac{1086766}{945} \zeta _2^3 \nonumber\\
    &\;\;-\mfrac{651151}{405} \zeta _5+\mfrac{92429}{162} \zeta _3 \zeta _2+\mfrac{1875703}{4860}\zeta _2^2+\mfrac{7693631}{1458}\zeta _3 \nonumber\\
    &\;\;+\mfrac{196729}{72} \zeta _2-\mfrac{749534537}{104976}-\mfrac{10}{3}\mathcal{H}\Big]+ \CS{N_f\text{\small{-terms}}},
\\[2ex]
\label{eq:ffneq4}
    &\FF{\bar{\mathcal{F}}_4^{\mathcal{N} = 4}(\epsilon)\Big|_{1/\epsilon}} = \mfrac{541619}{126}\zeta_7-\mfrac{15529}{45} \zeta_5 \zeta_2+\mfrac{39067}{135} \zeta_3 \zeta_2^2 \nonumber\\
    &\;\; +\CS{\frac{1}{N_c^2}}\Big[-\mfrac{7443}{4} \zeta_7 -\mint{492} \zeta_5 \zeta_2+\mfrac{1638}{5} \zeta_3 \zeta_2^2-\mint{1200}\zeta_3^2 \nonumber\\
    &\;\;-\mfrac{892}{7}\zeta_2^3+\mint{3000}\zeta_5+\mint{720}\zeta_3 \zeta_2 -\mint{36}\zeta_2^2-\mint{1080}\zeta_3
    +\mint{120}\mathcal{H}\Big].
\end{align}
To the best of our knowledge, we provide exact and unconditional results for the $1/\epsilon$ pole of the $C_F^4$ color structure, and therefore massless Quantum Electrodynamics, for the first time.

For the collinear anomalous dimensions of QCD and $\mathcal{N}=4$ SYM we obtain
\begin{align}
\label{eq:colq}
\phantom{.}&\hspace{-1.7ex}\FF{\gamma_4^q} =
    \CS{\frac{d_A^{abcd}d_F^{abcd}}{N_F}}\Big[\mint{9924}\zeta _7+\mint{2624} \zeta _5 \zeta _2-\mfrac{8736}{5} \zeta _3 \zeta _2^2
    +\mfrac{15856}{3} \zeta _3^2
    \nonumber\\
    & +\mfrac{241888}{315} \zeta _2^3 -\mfrac{145840}{9} \zeta _5-\mint{5632} \zeta _3 \zeta _2+\mfrac{3104}{15} \zeta _2^2 +\mfrac{44032}{9} \zeta _3  \nonumber \\
    & -\mfrac{2176}{3} \zeta _2+\mint{192}-\mint{640} \mathcal{H}\Big]+ \CS{C_A^3 C_F}\Big[-\mfrac{18927}{2} \zeta _7  \nonumber\\
   & +\mfrac{248}{3} \zeta _5 \zeta
   _2+\mfrac{956}{5} \zeta _3 \zeta _2^2-\mfrac{11674}{9} \zeta _3^2-\mfrac{139592}{315} \zeta _2^3 +\mfrac{301166}{27} \zeta _5
   \nonumber\\
   & +\mfrac{25480}{9} \zeta _3 \zeta _2+\mfrac{179182}{135} \zeta _2^2-\mfrac{2159464}{243} \zeta_3+\mfrac{1062149}{729} \zeta _2 \nonumber \\
   &  +\mfrac{7179083}{26244}+\mfrac{560}{3}\mathcal{H}\Big]+ \CS{C_A^2 C_F^2}\Big[\mint{22050} \zeta _7-\mint{3008} \zeta _5 \zeta _2 \nonumber \\
   & -\mfrac{6128}{5} \zeta _3 \zeta _2^2 +\mfrac{196}{3} \zeta _3^2-\mfrac{26136}{35} \zeta _2^3-\mfrac{97292}{9}\zeta_5-\mfrac{21728}{9} \zeta _3 \zeta _2 \nonumber \\
   & -\mfrac{44792 }{27}\zeta_2^2+\mfrac{375964}{27} \zeta _3 -\mfrac{93542}{27}\zeta _2+\mfrac{29639}{18}-\mint{480} \mathcal{H}\Big] \nonumber \\
   & + \CS{C_A C_F^3}\Big[-\mint{25060} \zeta _7+\mint{3328} \zeta _5 \zeta _2+\mfrac{4512}{5} \zeta _3 \zeta _2^2+\mint{3240}\zeta _3^2 \nonumber\\
   & +\mfrac{527336}{315} \zeta _2^3+\mint{6048} \zeta _5+\mfrac{1784}{3} \zeta _3 \zeta _2+\mfrac{8188}{5} \zeta_2^2-\mint{9400} \zeta _3 \nonumber \\
   & +\mint{2334} \zeta _2-\mfrac{2085}{2}+\mint{320} \mathcal{H}\Big] + \CS{C_F^4}\Big[\mint{11760} \zeta _7-\mint{768} \zeta _5 \zeta _2\nonumber \\
   & +\mfrac{256}{5} \zeta _3 \zeta _2^2-\mint{2304} \zeta_3^2-\mfrac{33776}{35} \zeta _2^3-\mint{5040} \zeta _5-\mint{240} \zeta _3 \zeta _2 \nonumber \\
   & -\mfrac{1368}{5} \zeta_2^2+\mint{4008} \zeta _3-\mint{900} \zeta _2+\mfrac{4873}{12}\Big]+\CS{N_f\text{\small{-terms}}},
\\[2ex]
&\hspace{-1.7ex}\FF{\gamma_4^g} =
    \CS{\frac{d_A^{abcd}d_A^{abcd}}{N_A}}\Big[\mint{9924} \zeta _7+\mint{2624} \zeta _5 \zeta _2-\mfrac{8736}{5} \zeta _3 \zeta _2^2 +\mfrac{15856}{3}\zeta _3^2 \nonumber \\
    &+\mfrac{253856}{315} \zeta _2^3-\mfrac{141280}{9} \zeta _5-\mint{6176} \zeta _3 \zeta _2+\mfrac{1072}{15} \zeta _2^2 +\mfrac{39328}{9} \zeta _3 \nonumber \\
    &+\mint{64} \zeta _2+\mfrac{128}{9}-\mint{640}\mathcal{H}\Big]+ \CS{C_A^4}\Big[-\mfrac{1427}{2} \zeta _7 -\mfrac{1096}{3} \zeta _5 \zeta_2 \nonumber\\
    &-\mfrac{404}{5} \zeta _3 \zeta _2^2-\mfrac{2686}{9} \zeta _3^2-\mfrac{100208}{135} \zeta _2^3+\mfrac{37232}{27} \zeta _5 +\mfrac{2068}{3} \zeta _3 \zeta _2 \nonumber \\
    &+\mfrac{248368}{135} \zeta_2^2-\mfrac{21940}{243} \zeta_3 -\mfrac{1051411}{729} \zeta_2 +\mfrac{10672040}{6561} +\mfrac{80}{3}\mathcal{H}\Big] \nonumber \\
    &+ \CS{N_f\text{\small{-terms}}},
\\[2ex]
\label{eq:colneq4}
    &\hspace{-1.7ex}\FF{\gamma_4^{\mathcal{N} = 4}} = -\mint{300} \zeta _7-\mint{256} \zeta _5 \zeta _2-\mfrac{768}{5} \zeta _3 \zeta _2^2+\CS{\frac{1}{N_c^2}}\Big[\mint{14886} \zeta _7 \nonumber\\
    &+\mint{3936} \zeta _5 \zeta _2 -\mfrac{13104}{5} \zeta _3 \zeta _2^2+\mint{9600} \zeta _3^2+\mfrac{7136}{7}\zeta _2^3 -\mint{24000} \zeta_5 \nonumber\\
    &-\mint{5760}\zeta _3 \zeta _2+\mint{288} \zeta _2^2 +\mint{8640} \zeta _3-\mint{960} \mathcal{H}\Big],
\end{align}
where we used the Supplementary Material of \cite{vonManteuffel:2020vjv} to extract the resummation functions in Eq.\ \eqref{eq:colqgdef} from our results for the poles of the form factors.
To the best of our knowledge, our planar-color $\mathcal{N} = 4$ SYM calculation is the first to independently confirm the analytic analysis of \cite{Dixon:2017nat}.

\section{Numerical results}
\label{sec:numresults}
In this section, we compile the explicit numerical results for the $1/\epsilon$ poles of the form factors and collinear anomalous dimensions, which we obtain using the numerical approximation for $\mathcal{H}$ \eqref{eq:num}.
We provide twelve digits for the color structures where exact analytic results are available ({\it i.e.} from \cite{vonManteuffel:2020vjv} for the $N_f$-dependent terms).
For the form factors we have
\begin{align}
\label{eq:numFq}
    \FF{\bar{\mathcal{F}}_4^q(\epsilon)\Big|_{1/\epsilon}} &\approx
    \CS{\frac{d_A^{abcd}d_F^{abcd}}{N_F}}\big[\mint{274.4588169341} \pm \mint{5.6} \cdot \mint{10}^{-9}\big] \nonumber\\
    &\quad + \CS{C_A^3 C_F}\big[-\mint{13274.5995371593} \pm \mint{1.6}\cdot \mint{10}^{-9}\big]\nonumber \\
    &\quad + \CS{C_A^2 C_F^2}\big[\mint{19661.7351772000} \pm \mint{4.2}\cdot \mint{10}^{-9}\big]\nonumber\\
    &\quad +\CS{C_A C_F^3}\big[-\mint{1602.0556057677} \pm \mint{2.8}\cdot \mint{10}^{-9}\big]\nonumber \\
    &\quad +\CS{C_F^4}\big[-\mint{2212.79784915}\myldots \big]
    \nonumber\\
    &\quad +\CS{N_f \frac{d_F^{abcd}d_F^{abcd}}{N_F}} \big[\mint{53.1274437988}\myldots\big] \nonumber\\
    &\quad +\CS{N_f C_A^2 C_F}\big[\mint{10203.6391859}\myldots\big] \nonumber\\
    &\quad +\CS{N_f C_A C_F^2}\big[-\mint{12551.5075480}\myldots\big] \nonumber \\
    &\quad +\CS{N_f C_F^3} \big[\mint{2095.46596925}\myldots\big]\nonumber \\
    &\quad +\CS{N_{q \gamma} C_A \frac{d_F^{abc}d_F^{abc}}{N_F}} \big[-\mint{235.381562912}\myldots\big] \nonumber \\
    &\quad +\CS{N_{q \gamma} C_F \frac{d_F^{abc}d_F^{abc}}{N_F}} \big[-\mint{243.738662819}\myldots\big] \nonumber\\
    &\quad +\CS{N_f^2 C_A C_F} \big[-\mint{2304.68219272}\myldots\big] \nonumber \\
    &\quad +\CS{N_f^2 C_F^2}  \big[\mint{1604.85115658}\myldots\big] \nonumber\\
    &\quad +\CS{N_{q \gamma} N_f \frac{d_F^{abc}d_F^{abc}}{N_F}} \big[\mint{42.7966478022}\myldots\big] \nonumber\\ 
    &\quad + \CS{N_f^3 C_F} \big[\mint{158.065537245}\myldots\big],
\\[2ex]
\label{eq:numFg}
    \FF{\bar{\mathcal{F}}_4^g(\epsilon)\Big|_{1/\epsilon}} &\approx
    \CS{\frac{d_A^{abcd}d_A^{abcd}}{N_A}}\big[\mint{579.5738867755} \pm \mint{5.6} \cdot \mint{10}^{-9}\big] \nonumber \\
    &\quad + \CS{C_A^4}\big[-\mint{1081.02280574667} \pm \mint{2.3} \cdot \mint{10}^{-10}\big]\nonumber\\
    &\quad +\CS{N_f \frac{d_A^{abcd}d_F^{abcd}}{N_A}} \big[-\mint{604.701004352}\myldots\big]  \nonumber\\
    &\quad +\CS{N_f C_A^3} \big[-\mint{532.481107793}\myldots\big] \nonumber\\
    &\quad +\CS{N_f C_A^2 C_F} \big[-\mint{650.895247054}\myldots\big] \nonumber\\
    &\quad +\CS{N_f C_A C_F^2} \big[\mint{14.9978706950}\myldots\big] \nonumber\\
    &\quad +\CS{N_f C_F^3} \big[ \mint{17.25}  \big] \nonumber\\
    &\quad +\CS{N_f^2\frac{d_F^{abcd}d_F^{abcd}}{N_A}} \big[\mint{95.1966169377}\myldots\big] \nonumber\\
    &\quad +\CS{N_f^2 C_A^2}\big[\mint{1574.06171919}\myldots\big] \nonumber\\
    &\quad +\CS{N_f^2 C_A C_F}\big[\mint{282.052204632}\myldots\big] \nonumber\\
    &\quad +\CS{N_f^2 C_F^2} \big[\mint{23.4647858335}\myldots\big] \nonumber\\
    &\quad +\CS{N_f^3 C_A}\big[-\mint{261.955705460}\myldots\big]\nonumber\\
    &\quad +\CS{N_f^3 C_F} \big[-\mint{12.7425646335}\myldots\big],
\\[2ex]
    \FF{\bar{\mathcal{F}}_4^{\mathcal{N} = 4}(\epsilon)\Big|_{1/\epsilon}} &\approx \mint{4687.07846404}\myldots \nonumber\\
    &\quad + \CS{\frac{1}{N_c^2}}\big[-\mint{896.4243270825} \pm \mint{8.4} \cdot \mint{10}^{-9}\big].
\end{align} 
Comparing Eq.\ \eqref{eq:numFq} and Eq.\ \eqref{eq:numFg} to, respectively, Eq.\ (3.27) of \cite{Das:2019btv} and Eq.\ (11) of \cite{Das:2020adl}, we find that our results agree completely to within their given error estimates. We note that our Eqs.\  \eqref{eq:numFq} and \eqref{eq:numFg} significantly improve upon the QCD results of \cite{Das:2019btv,Das:2020adl}, as the numerical approximations provided therein are, depending on the color structure, accurate to, at best, six significant digits ({\it e.g.} $C_A^2 C_F^2$) and, at worst, one significant digit ({\it e.g.} $d^{abcd}_A d^{abcd}_A/N_A$).
For the collinear anomalous dimensions we have
\begin{align}
    \FF{\gamma_4^q} &\approx 
    \CS{\frac{d_A^{abcd}d_F^{abcd}}{N_F}}\big[-\mint{2195.670535473} \pm \mint{4.5} \cdot \mint{10}^{-8}\big] \nonumber\\
    &\quad + \CS{C_A^3 C_F}\big[-\mint{13.809312037} \pm \mint{1.3}\cdot \mint{10}^{-8}\big]\nonumber \\
    &\quad + \CS{C_A^2 C_F^2}\big[\mint{2438.569338812} \pm \mint{3.3} \cdot \mint{10}^{-8}\big] \nonumber\\
    &\quad + \CS{C_A C_F^3}\big[-\mint{1373.764650948} \pm \mint{2.2} \cdot \mint{10}^{-8}\big]\nonumber \\
    &\quad +\CS{C_F^4} \big[\mint{392.899478384}\myldots\big] \nonumber\\
    &\quad +\CS{N_f \frac{d_F^{abcd}d_F^{abcd}}{N_F}} \big[-\mint{425.019550390}\myldots\big]\nonumber\\
    &\quad +\CS{N_f C_A^2 C_F} \big[-\mint{274.147360589}\myldots\big]\nonumber\\
    &\quad +\CS{N_f C_A C_F^2} \big[-\mint{912.844845636}\myldots\big]\nonumber\\
    &\quad +\CS{N_f C_F^3} \big[\mint{151.933788877}\myldots\big]\nonumber\\
    &\quad +\CS{N_f^2 C_A C_F} \big[\mint{109.081415293}\myldots\big]\nonumber\\
    &\quad +\CS{N_f^2 C_F^2} \big[-\mint{12.5342425083}\myldots\big]\nonumber\\
    &\quad +\CS{N_f^3 C_F} \big[\mint{4.88682798281}\myldots\big]\,,
\\[2ex]
    \FF{\gamma_4^g} &\approx
    \CS{\frac{d_A^{abcd}d_A^{abcd}}{N_A}}\big[-\mint{2451.040712450} \pm \mint{4.5} \cdot \mint{10}^{-8}\big] \nonumber\\
    &\quad +\CS{C_A^4}\big[\mint{1557.4287417889} \pm \mint{1.9} \cdot \mint{10}^{-9}\big]\nonumber\\
    &\quad +\CS{N_f \frac{d_A^{abcd}d_F^{abcd}}{N_A}} \big[-\mint{41.2080190194}\myldots\big] \nonumber\\
    &\quad +\CS{N_f C_A^3} \big[-\mint{1033.98729659}\myldots\big] \nonumber\\
    &\quad +\CS{N_f C_A^2 C_F} \big[-\mint{57.9377499658}\myldots\big] \nonumber\\
    &\quad +\CS{N_f C_A C_F^2} \big[-\mint{100.315097910}\myldots\big] \nonumber\\
    &\quad +\CS{N_f C_F^3} \big[\mint{46}\big] \nonumber\\
    &\quad +\CS{N_f^2\frac{d_F^{abcd}d_F^{abcd}}{N_A}} \big[\mint{253.857645167}\myldots\big] \nonumber\\
    &\quad +\CS{N_f^2 C_A^2} \big[\mint{70.7744401902}\myldots\big] \nonumber\\
    &\quad +\CS{N_f^2 C_A C_F} \big[\mint{73.9372035966}\myldots\big] \nonumber\\
    &\quad +\CS{N_f^2 C_F^2} \big[-\mint{21.9767440643}\myldots\big] \nonumber\\
    &\quad +\CS{N_f^3 C_A} \big[\mint{0.405507202650}\myldots\big] \nonumber\\
    &\quad + \CS{N_f^3 C_F} \big[\mint{1.26748971193}\myldots\big],
\\[2ex]
\label{eq:colneq4num}
    \FF{\gamma_4^{\mathcal{N} = 4}} &\approx -\mint{1238.74771725}\myldots \nonumber\\
    &\quad + \CS{\frac{1}{N_c^2}}\big[\mint{7171.394616660} \pm \mint{6.7} \cdot \mint{10}^{-8}\big]\, .
\end{align}
The errors provided in the above equations are all statistical errors from the {\tt pySecDec} evaluation of $\mathcal{H}$. For the non-planar-color part of the $\mathcal{N} = 4$ SYM collinear anomalous dimension, Eq.\ \eqref{eq:colneq4num} is consistent with the result of \cite{Boels:2017ftb} within their provided error estimate, but it is a vast improvement over Eq.\ (5.8) of \cite{Boels:2017ftb} as the latter gives essentially only an order of magnitude estimate for this quantity. Besides our comparisons to the existing literature, we also carried out numerical cross-checks on many of the most complicated four-loop form factor master integral expansion coefficients using either {\tt pySecDec} or {\tt FIESTA 4} \cite{Smirnov:2015mct}.

\section{Lifting numerical data to analytic expressions}
\label{sec:pslqanalysis}

In this section, we analyze the unknown analytic form of $\mathcal{H}$.
First, assuming that the $\epsilon$-expansion coefficients of $I_{6}^{(26)}$ and $I_{7}^{(26)}$ are multiple zeta values of uniform weight implies that $\mathcal{H}$ has the form
\begin{align}
\label{eq:Hreduced}
    \mathcal{H} &= a \zeta _7 + b \zeta _5 \zeta _2 + c \zeta _3 \zeta _2^2 +\mint{10} \zeta _3^2 + \mfrac{223}{210}\zeta _2^3-\mint{25} \zeta _5 \nonumber \\
    &\quad - \mint{6} \zeta _3 \zeta _2+\mfrac{3}{10}\zeta _2^2+\mint{9} \zeta _3
\end{align}
for some rational numbers $a$, $b$ and $c$.
This constraint will be central to the following discussion.

Let us consider the unknown part of Eq.\ \eqref{eq:Hreduced},
\begin{equation}
\label{eq:testnum}
    a \zeta _7 + b \zeta _5 \zeta _2 + c \zeta _3 \zeta _2^2 \approx
    \mint{6.279370861144} \pm \mint{7.0} \cdot \mint{10}^{-11} \,,
\end{equation}
and fit the constants $a$, $b$, and $c$ using the PSLQ algorithm. Our first task will be to assess the uncertainty of our high-precision run of {\tt pySecDec} by comparing approximation \eqref{eq:num} to a preliminary run of the program at a somewhat lower precision,
\begin{equation}
\label{eq:oldnum}
    \mathcal{H} \approx
    -\mint{0.7015802399} \pm \mint{2.75} \cdot \mint{10}^{-8} \,.
\end{equation}
Subtracting the central value of \eqref{eq:num} from the central value of \eqref{eq:oldnum} and dividing by the uncertainty of \eqref{eq:oldnum}, we find a ratio of 1.2. This indicates that the initial estimate produced by {\tt pySecDec} was a bit too large but that the uncertainty estimate produced by the program seems trustworthy. 
We will proceed under the assumption that the given statistical error faithfully represents the actual uncertainty of the approximation, excluding in particular the logical possibility of a substantial but hidden systematic shift.
In particular, we consider 11 digits in \eqref{eq:testnum} to be significant and 10 digits to be safe.

In order to obtain a rough estimate for the required number of digits for a successful fit, we considered the complexity of rational numbers appearing in a sample set of analytically known integrals from other topologies.
Here, we selected all 318 integrals whose leading term in the $\epsilon$ expansion involves weight 7 zeta values, and which might therefore be similar to the unknown integral.
We find that, on average, 10 digits were required to successfully reconstruct the rational coefficients of weight 7 zeta values.
It therefore seems possible that the right-hand side of \eqref{eq:testnum} could suffice to fit the rational constants on the left-hand side.

While retaining ten digits of approximation \eqref{eq:testnum} is not quite good enough, retaining eleven digits results in the very promising fit:
\begin{equation}
\label{eq:fit}
    a = \mfrac{161}{16},\quad b = \mfrac{5}{2},\quad c = -\mfrac{5}{2}\,.
\end{equation}
We now attempt to quantify whether these putative values for $a$, $b$, and $c$ are reasonable or not using statistical arguments and numerical extrapolation. This analysis is of considerable importance, as our fit stretches approximation \eqref{eq:testnum} to its limit. As a start, note that subtracting the presumptive exact value obtained from fit \eqref{eq:fit} from the central value of \eqref{eq:num} and dividing by its uncertainty, we find a quite reasonable ratio of 1.1.

Our PSLQ fit \eqref{eq:fit} seems plausible in the sense, that the rational numbers involve relatively small integers and are rather similar in structure to what we observe for the rational prefactors of $\{\zeta_7,\zeta_5 \zeta_2,\zeta_3 \zeta_2^2\}$ in our sample of 318 superficially-similar leading-order integral expansion coefficients.\footnote{
By experimenting with our sample expressions, we found that fewer significant digits are generally required for a fit in terms of $\{\zeta_7,\zeta_5 \zeta_2,\zeta_4 \zeta_3\}$, which is therefore the default ansatz for our actual PSLQ runs.
}
Indeed, we observe that for all 318 elements of our sample, the denominators of the rational prefactors in front of $\{\zeta_7,\zeta_5 \zeta_2,\zeta_3 \zeta_2^2\}$ are {\it never} larger than $\{75,12,240\}$ and the prime factors which appear in the prime factorizations of the denominators are {\it never} larger than $\{5,3,5\}$.
Moreover, the number $161/16$ we found for $a$ is actually a quite typical one with respect to our 318 samples of $\zeta_7$ coefficients; its numerator appears in front of $\zeta_7$ 10 times in our sample and its denominator is actually the most common one in our sample, appearing 111 times as the denominator of the rational prefactor of $\zeta_7$.

In order to explore the robustness of the fit \eqref{eq:fit}, we systematically extrapolated the ten significant digits of approximation \eqref{eq:testnum}. To be precise, we continue its decimal expansion in all possible ways out to a maximum of fifteen significant digits.\footnote{It turns out that {\it all} weight seven terms of our 318 sample expansion coefficients can be reconstructed by running the PSLQ algorithm with fifteen significant digit input precision.} After each digit is added by our trial code, a PSLQ fit is attempted and then judged according to the simplicity and similarity criteria of the previous paragraph. We find it remarkable that, all the way out to fifteen decimal digits, no other possible PSLQ fit looks nearly as natural as \eqref{eq:fit} above. For this extrapolation analysis, we only considered decimal expansions lying within plus or minus two times the reported statistical uncertainty on $\mathcal{H}$. In an abundance of caution, we repeated our analysis with only nine digits of approximation \eqref{eq:testnum} and found nothing different for extrapolations out to fourteen decimal digits lying within plus or minus five times the reported statistical uncertainty on $\mathcal{H}$.

Inserting our fit for $\mathcal{H}$ into \eqref{eq:colneq4} we find
for the collinear anomalous dimension in $\mathcal{N}=4$ SYM
\begin{align}
    \label{eq:finalcol4neq4}
    \FF{\gamma_4^{\mathcal{N} = 4}} &= -\mint{300} \zeta _7-\mint{256} \zeta _5 \zeta _2 -\mint{384} \zeta _4\zeta _3 \nonumber\\
    &\quad +\CS{\frac{1}{N_c^2}}\big[\mint{5226} \zeta _7+\mint{1536} \zeta _5 \zeta _2-\mint{552} \zeta _4\zeta _3\big].
\end{align}
We see that Eq.\ \eqref{eq:finalcol4neq4} is built entirely out of integer linear combinations of weight seven zeta values if one chooses the basis element $\zeta_4 \zeta_3$ instead of $\zeta_3 \zeta_2^2$.

\section{Conclusion}
\label{sec:summary}
In this work, we obtained precise numerical approximations for the four-loop collinear anomalous dimensions of QCD and $\mathcal{N} = 4$ SYM as well as
conjectures for the full analytic results in terms of zeta values.
Our experiments with {\tt pySecDec} suggest that it should be possible to numerically evaluate the finite parts of the QCD form factors to sufficiently high precision for phenomenological purposes.
\vspace{.3 cm}

\noindent{\em Acknowledgments:}
BA and AvM are supported in part by the National Science Foundation through Grant 2013859.
EP is funded as a Royal Society University Research Fellow through grant {URF{\textbackslash}R1{\textbackslash}201473}, and previously by All Souls College Oxford. 
We acknowledge the High Performance Computing Center at Michigan State University
for computing resources and thank the team for their help and support.
We also thank the Mathematical Institute at the University of Oxford for computing resources.

\bibliography{ff4lcollinear}
\end{multicols}
\end{document}